\documentclass[epj]{svjour}
 
\usepackage{epsfig}
\usepackage{amssymb,amsmath,amsfonts,dsfont}
\usepackage{graphicx}
\usepackage{ulem}
\usepackage{widetext}
\usepackage{color}

\newcommand{\bE}{\mathbf{E}}

\newcommand{\lmax}{\ell_\mathrm{max}}

\newcommand{\brho}{\boldsymbol{\rho}}

\newcommand{\bH}{\mathbf{H}}
\newcommand{\bMM}{\mathbf{M}}

\newcommand{\calD}{\mathcal{D}}

\newcommand{\calS}{\mathcal{S}}

\newcommand{\bN}{\mathbf{N}}

\newcommand{\br}{\mathbf{r}}

\newcommand{\bp}{\mathbf{p}}
\newcommand{\bu}{\mathbf{u}}

\def\d{\mathrm{d}}
\def\eps{\varepsilon}

\def\Im{\mathrm{Im}}

\def\Id{\mathrm{Id}}

\def\inc{\mathrm{inc}}

\def\free{\mathrm{free}}
\def\exit{\mathrm{exit}}

\def\sph{\mathrm{sph}}
\def\sca{\mathrm{sca}}
\def\exc{\mathrm{exc}}

\def\iin{\mathrm{in}}
\def\out{\mathrm{out}}

\newcommand{\Ylm}{Y_{\ell,m}}
\newcommand{\Alm}{A_{\ell,m}}
\newcommand{\Blm}{B_{\ell,m}}
\newcommand{\Plm}{P_{\ell}^{m}}

\renewcommand{\emph}{\textit}

\newcommand{\ampl}{P_0}
\newcommand\mydots{\hbox to 1em{.\hss.\hss.}}

\begin{document}

\title{Scattering properties of collective dipolar systems}

\date{\today}

\author{A. Canaguier-Durand\inst{1}  \and A. Lambrecht\inst{1} \and S. Reynaud\inst{1}}

\institute{                    
  \inst{1} Laboratoire Kastler Brossel, Sorbonne Universit\'{e}, CNRS, ENS-PSL University, Coll\`{e}ge de France, Paris, France.
}


\abstract{
We present a theoretical treatment of light scattering by an ensemble of $N$ dipoles, taking into account recurrent multiple scattering. We study the intrinsic optical properties of collective dipolar systems without specifying a particular illumination condition. We apply this formalism to study the collective absorption modes for an ensemble of small nanoparticles and then to derive collective radiative corrections due to mutual interactions in dipolar ensembles, a topic of major importance in the development of collective nanophotonic systems or atomic networks used as optical clocks.}

\maketitle

\section*{Introduction}
Collective phenomena are involved in many innovative nanophotonic systems and they take benefit of the richness of interactions between emitters \cite{scheibner2007superradiance,schwartz2011reversible,george2015ultra,goban2015superradiance,bromley2016collective,bettles2016enhanced,zhu2016light}. In atomic ensembles in particular, collective coupling can be a powerful mean to modulate the emission properties, by controlling superradiant emission \cite{dicke1954coherence,gross1982superradiance,goban2015superradiance,li2016controlling,solano2017super} or generating a spectral shift \cite{rohlsberger2010collective,keaveney2012cooperative}. 
Yet it could also represent a genuine obstacle, as in the case of optical atomic clocks engineering. Atomic networks can indeed combine a large number of trapped atoms with ultra-narrow transitions to provide today's most precise clocks \cite{le2013experimental,bloom2014optical,ludlow2015optical,marti2018imaging} while interactions can cause frequency shifts \cite{chang2004controlling,hazlett2013s} that it is crucial to control in a metrological context. 

In this article we present a theoretical framework that enables to calculate the scattering matrix for an ensemble of dipoles. 
This framework is built up on the scattering formalism which turned out to be a very powerful tool to describe interaction of light with one or two scatterers. In particular, the dispersive interaction energy between two atoms or two nanoparticles can be directly derived from the scattering matrix \cite{jaekel1991casimir,lambrecht2006casimir}. Meanwhile, we use theoretical and numerical methods which have been developed to tackle multiple scattering,  in particular the Coupled Dipoles Model (CDM) which is very well suited to account for recurrent scattering taking place in a collective or dense system \cite{lax1952multiple,purcell1973scattering,chaumet2004coupled}. 

The new framework merges the two approaches and describes light scattering by an ensemble of dipoles. It calculates a collective scattering matrix including multiple scattering at all orders. Moreover it describes the intrinsic optical properties of the ensemble, without any specific choice of illumination or emission conditions. We will illustrate the advantages of this model, by studying first the collective absorption properties of an ensemble of dielectric nanospheres, and by computing then the collective radiative corrections to energy levels due to mutual interactions in an ensemble of dipoles.

\section{The basic approach}  A polarizable dipole located at $\br_0$ and submitted to an incoming harmonic field $\bE_\inc(\br)$ of pulsation $\omega=ck$ acquires a dipolar momentum 
\begin{align*}
\bp_0 = \eps_0 \alpha_0(\omega) \bE_\inc(\br_0)~,
\end{align*}
where its polarizability $\alpha_0$ has the dimension of a volume. It then generates a scattered field which can be expressed by using the Green tensor (see Appendix \ref{app:green} for explicit expression) as 
\begin{align*}
\bE_\sca(\br) = \mu_0 \omega^2 \mathbf{G}_0(\br,\br_0,\omega) \cdot \bp_0~.
\end{align*}
In a collective system of $N$ interacting dipoles, each dipole additionally receives the field scattered by other dipoles. This is  treated exactly by replacing for each dipole the incoming field by an exciting field accounting for the emission of other dipoles
\begin{align*}
\bE_{\exc,i}(\br) = \bE_\inc(\br) + \mu_0 \omega^2 \sum_{i'\neq i} \mathbf{G}_0(\br,\br_{i'},\omega) \cdot \bp_{i'} ~. 
\end{align*}

 Writing the exciting fields $\bE_i = \bE_{\exc,i}(\br_i)$ as components of a column vector form, it is easy to express them from the incident fields by inverting a linear system 
\begin{align}\label{Ei_wrt_inc_ri}
\begin{pmatrix} \vdots \\ \bE_i \\ \vdots  \end{pmatrix} = \left[ I - X \right]^{-1}  \begin{pmatrix} \vdots \\ \bE_\inc(\br_i) \\ \vdots  \end{pmatrix}   ~,
\end{align}
where  $X$ is a structure matrix of size $3N$ by $3N$, whose blocks $\left(X_{ii'} \right)_{i\neq i'}$ are 3-by-3 matrices containing the Green tensor that describes the propagation between the dipoles in vacuum (the diagonal blocks $X_{ii}$ are zero)
\begin{align}\label{Ei_wrt_inc_ri_X}
X_{ii'}=   \left( k^3 \alpha_{i'}  \frac{\mathbf{G}_0(\br_i,\br_{i'},\omega)}{k} \right)_{i\neq i'} ~ .
\end{align}

The field scattered by the ensemble of dipoles is then written by linearity as
\begin{align}\label{Esca_wrt_Ei}
\bE_\sca(\br) 
= \left[ \mydots , \left( k^3 \alpha_i\right) \frac{\mathbf{G}_0(\br,\br_i,\omega)}{k} , \mydots \right]  \begin{pmatrix} \vdots \\ \bE_i \\ \vdots  \end{pmatrix}  ~,
\end{align}
where the multiplying operator on the left is made of $N$ blocks  of size $3$ by $3$. Combining (\ref{Ei_wrt_inc_ri}-\ref{Esca_wrt_Ei}), one obtains an explicit expression of scattered fields from incident ones. 

We now combine  steps (\ref{Ei_wrt_inc_ri}-\ref{Esca_wrt_Ei}) of the CDM approach with the scattering of electromagnetic modes. 
Any harmonic electric field is decomposed into spherical modes as
\begin{align}\label{def_EH}
\bE(\br) = &\frac{\ampl}{\sqrt{\eps_0}} \sum_{\ell=1}^{\infty} \sum_{m=-\ell}^{\ell} A_{\ell,m}^{(1)} \bN_{\ell,m}^{(1)}(\br) + A_{\ell,m}^{(2)} \bN_{\ell,m}^{(2)}(\br) \nonumber \\
 & ~  ~  ~  ~ + \imath B_{\ell,m}^{(1)} \bMM_{\ell,m}^{(1)}(\br) + \imath B_{\ell,m}^{(2)} \bMM_{\ell,m}^{(2)} (\br)~,
\end{align}
where the prefactor $\ampl$ is a normalizing amplitude (see Appendix \ref{app:spherical} for explicit expressions and more practical details on this vector spherical modes decomposition). Coefficients $\Alm^{(1)}$, $\Blm^{(1)}$ correspond to modes going outwards and $\Alm^{(2)}$, $\Blm^{(2)}$ to modes coming inwards. 
This enables to define the linear subspace of purely outgoing fields $E_\exit$ for which the coefficients are all zero for inward modes 
\begin{align*}
&A_{\ell,m}^{(2)}= B_{\ell,m}^{(2)}=0 ~ ~  \mathrm{for ~ all} ~(\ell,m) ~.
\end{align*}
For such a field we introduce the compact notation 
\begin{align*}
\phi_\exit=(\Alm^{(1)},\Blm^{(1)})^T ~.
\end{align*}
We also introduce the linear subspace of free fields $E_\free$  as the set of all fields whose coefficients obey 
\begin{align*}
&A_{\ell,m}^{(1)}=A_{\ell,m}^{(2)} &B_{\ell,m}^{(1)}=B_{\ell,m}^{(2)} ~~\mathrm{for ~ all} ~(\ell,m)
~.
\end{align*}
For those fields, we introduce the coefficients 
\begin{align*}
&\Alm^{(j)}=2\Alm^{(1)}= 2\Alm^{(2)}~, &\Blm^{(j)}=2 \Blm^{(1)}=2\Blm^{(2)}~,
\end{align*}
 which are naturally multiplied by the modes 
\begin{align*}
\bMM_{\ell,m}^{(j)} (\br) = \left[ \bMM_{\ell,m}^{(1)} (\br) + \bMM_{\ell,m}^{(2)} (\br) \right] / 2 ~, \\
\bN_{\ell,m}^{(j)} (\br) = \left[ \bN_{\ell,m}^{(1)} (\br) + \bN_{\ell,m}^{(2)} (\br) \right] / 2 ~,
\end{align*}
whose expressions are the same as $\bMM^{(q)}_{\ell,m}$ and $\bN^{(q)}_{\ell,m}$ with Riccati-Hankel functions $\xi^{(q)}_\ell$ replaced by Riccati-Bessel functions $\psi_\ell$. As a consequence these fields are finite at the origin and are solutions of the Maxwell equation in vacuum. For this subspace we also introduce a compact notation as \begin{align*}
\phi_\free=(\Alm^{(j)},\Blm^{(j)})^T
~.
\end{align*}
 
In the following, for practical implementations a finite number $N_\sph$ of spherical modes will be considered, with $\ell \leq \ell_\mathrm{max}$ and $|m| \leq \ell$ which yields $N_\sph = 2\lmax(\lmax+2)$. Globally, $\ell_\mathrm{max}$ must be larger for collective systems where dipoles are closer to each other and in practice it must be increased until the result reaches convergence. For the cases considered in this article, numerical computations have been performed with values of $\ell_\mathrm{max}$ from 8 ($N_\sph=160$) up to 24 ($N_\sph=1248$).

\section{Two descriptions of scattering} 
The diffusion of an incoming field on a single object can be described by two different but equivalent methods. 
In the first one, commonly used in classical optical approach, one considers the \emph{incident} and \emph{scattered} field, which together add up to be the total field with 
\begin{align*}
&\phi_\inc \in E_\free , &\phi_\sca \in E_\exit
~.
\end{align*}
 In the second one, which is related to the scattering formalism and more common in quantum optical approach, the \emph{in} and \emph{out} fields are connected through interaction with the system. These fields must be solutions in the absence of the system and are hence free fields
\begin{align*}
\phi_\iin, \phi_\out \in E_\free
~.
\end{align*}

These two approaches define two distinct transformation matrices that we will denote $D$ (for \emph{diffusion} matrix) and $S$ (for \emph{scattering} matrix) as
\begin{align}\label{DS_isolated}
&\phi_\sca = D \cdot \phi_\inc ~,
&\phi_\out = S \cdot \phi_\iin ~,
\end{align}
which have respective codomains $E_\exit$ and $E_\free$. 
These two matrices are related through the equality of the total field, using $\phi_\iin=\phi_\inc$ and $\phi_\out = 2\phi_\sca+\phi_\inc$ \cite{fernandez2017unified},
\begin{align}\label{DS_relation}
S = I + 2 D   ~.
\end{align} 
It must be kept in mind that this relation connects the coefficients $\Alm^{(j)},\Blm^{(j)}$ for $S$ in the left-hand side to the coefficients $\Alm^{(1)},\Blm^{(1)}$ for $D$ in the right-hand side.

The CDM, that is well adapted to describe the multiple scattering events occurring among the dipoles, uses incident and scattered fields $\phi_\inc$, $\phi_\sca$ and hence is naturally connected to the diffusion operator. As a consequence our strategy will be to first compute the diffusion operator for the collective system, that we will call $\mathcal{D}$ in the following, by translating eqs.~(\ref{Ei_wrt_inc_ri},\ref{Esca_wrt_Ei}) into spherical modes. The obtained results will then be rewritten in terms of free fields using (\ref{DS_relation}), in order to deduce the collective scattering operator $\mathcal{S}$.

\section{The collective scattering matrix} 
We start with eq.~(\ref{Ei_wrt_inc_ri}), for which one has to evaluate the incident field at the dipoles' positions $\left( \br_i \right)$ from its multipolar coefficients  expressed in a common  frame
\begin{align*}
\phi_\inc = (\Alm^{(j)},\Blm^{(j)})^T
~.
\end{align*}
To do so, we proceed in two steps: first we translate the incident field by $\br_i$ to obtain the coefficients in spherical modes when it is expressed with respect to $\br_i$ with the operator $T_{i0}$ that transforms a free field to another free field \cite{wittmann1988spherical} (see Appendix \ref{app:Tmatrix} for explicit expressions); then we use these spherical coefficients to evaluate the vector field at the new origin $\br_i$. This second step can be realized through multiplication by an $F$ matrix of size 3 by $N_\sph$ (See Appendix \ref{app:Fmatrix} for details):
\begin{align*}
&\frac{\bE(\mathbf{0})}{\ampl/\sqrt{\eps_0}} = \underbrace{\frac{1}{\sqrt{12\pi}} \begin{pmatrix} 
 \imath & 0 & -\imath & 0 & \mydots \\
 1 & 0 & 1 & 0 & \mydots \\
 0 & \sqrt{2} \imath & 0 & 0 & \mydots  \end{pmatrix}}_{F}
 \phi_\inc  ~.
\end{align*}
Here $\bE(\mathbf{0})$ is written in Cartesian coordinates and the only non-zero coefficients of the $F$ matrix correspond to $\Alm^{(j)}$ with $\ell=1$ and $m \in \{-1,0,1 \}$. 
Finally, coming back to eq.~(\ref{Ei_wrt_inc_ri}) one obtains the exciting fields from the incident field as 
\begin{align}\label{dip_part1}
\frac{1}{ \ampl/\sqrt{\eps_0}} \begin{pmatrix} \vdots \\ \bE_i  \\ \vdots  \end{pmatrix} = [I-X]^{-1} \begin{pmatrix} \vdots \\ F  T_{i0} \\ \vdots \end{pmatrix} \phi_\inc 
\end{align}
where the operator in parenthesis is a $3N$ by $N_\sph$ matrix.

We then consider eq.(\ref{Esca_wrt_Ei}), for which one has to express the coefficients $\phi_\sca = (\Alm^{(1)},\Blm^{(1)})^T$ of the scattered field. Again, we do this in two steps, first by expressing in spherical modes the field scattered by a dipole at the origin and then by operating a translation by $-\br_i$ for the obtained field. If one considers a single dipole  $\bp_0=\eps_0 \alpha_0 \bE_\inc(\mathbf{0})$ located at the origin, the coefficients of the produced scattered field $\phi_\sca^{0}$ are obtained as a multiplication by a $N_\sph$ by 3 matrix $Q$ (see Appendix \ref{app:Qmatrix} for details):
\begin{align*}
\phi_\sca^{0} =  k^3 \alpha_0 \underbrace{\frac{1}{\sqrt{12\pi}} \begin{pmatrix} 
1 & \imath  & 0 \\ 0 & 0 & \sqrt{2} \\ -1 & \imath & 0 \\ 
  \vspace{-0.1cm} 0 & 0 & 0 \\  \vspace{0.cm} \vdots & \vdots & \vdots  \end{pmatrix}}_{Q}
\frac{\bE_\inc(\mathbf{0})}{\ampl/\sqrt{\eps_0}}                                    ~.
\end{align*}
Non-zero components of $Q$ are relative to $\Alm^{(1)}$ coefficients with $\ell=1$ and $m \in \{-1,0,1 \}$, and $\bE_\inc(\mathbf{0})$ is written in cartesian coordinates. We have separated the dimensionless factor $k^3 \alpha_0$ as it  depends on the dipole polarizability, while $Q$ is a universal operator, related to $F$ by
\begin{align*}
Q=\imath F^\dagger~.
\end{align*}
 Then the scattered field can be translated by $(-\br_i)$ such that it represents a field generated by a dipole at $\br_i$ in the basis associated with the global reference point $\br_0$. The translation is realized by the operator $T_{0i}$ that transforms an outgoing field to another outgoing field \cite{wittmann1988spherical} (see Appendix \ref{app:Tmatrix} for explicit expressions). Finally doing this for each scatterer, we can express the total scattered field from the exciting fields as:
\begin{align}\label{dip_part2}
&\phi_\sca = \begin{pmatrix} \mydots & , ~(k^3 \alpha_i) T_{0i} Q ~ ,  & \mydots \end{pmatrix}   \frac{1}{\ampl/\sqrt{\eps_0}}  \begin{pmatrix} \vdots \\  \bE_i  \\ \vdots  \end{pmatrix}  ~, 
\end{align}
where the constructed matrix on the left has dimensions $N_\sph$ by $3N$. 

Combining results (\ref{dip_part1}) and (\ref{dip_part2}) , we obtain the matrix $\mathcal{D}$ of the total system as
\begin{align}\label{D_dip_result}
\mathcal{D}= \begin{pmatrix} \mydots & , ~ (k^3 \alpha_i) T_{0i} Q  , &  \mydots \end{pmatrix}  
 [I-X]^{-1}  \begin{pmatrix} \vdots \\ F T_{i0} \\ \vdots \end{pmatrix} 
\end{align}
and then the scattering matrix $\mathcal{S}$ by using (\ref{DS_relation}),
which is the main result of this article.
The operator $\mathcal{S}$ fully describes the collective scattering by the ensemble of $N$ dipoles without specifying a specific choice of incident field, thus characterizing the intrinsic scattering properties of the collective dipolar system. As discussed in the following, this formalism is able to determine the  proper scattering modes generated by collective interactions in such a system. Moreover, expanding the inverted operator into its Taylor series $(I+X+X^2+ \cdots)$ we get an expansion of the scattering matrix $\mathcal{S}= I+\mathcal{S}_1+\mathcal{S}_2+\cdots$ where $\mathcal{S}_k$ describes fields scattered $k$ times on various objects. This highlights the fact that this approach is non-perturbative as it includes multiple scattering at all orders.

\section{Unitarity and the optical theorem} 
Before studying applications of this formalism, we note that the unitarity of the scattering matrix is linked to the optical theorem that ensures energy conservation or equivalently probability conservation in a situation where frequency conversion is disregarded. 

For an isolated dipole located at the origin, formula (\ref{D_dip_result}) is simplified to 
\begin{align*}
D = (k^3 \alpha) QF = (\imath k^3 \alpha/6\pi) I_{(A,\ell=1)} ~,
\end{align*}
where $I_{(A,\ell=1)}$ is the identity operator for coefficients $A_{\ell,m}$ with $\ell=1,~m\in \{-1, 0,  1 \}$ and zero elsewhere. The condition of unitarity for the matrix $S$ of a single dipole can then be written explicitly as
\begin{align}\label{S_unitary_cond}
I-S^\dagger S = \frac{2k^3}{3\pi} \left[ \Im[\alpha] - \frac{k^3 |\alpha|^2}{6\pi} \right] I_{(A,\ell=1)} = \mathbf{0}~,
\end{align}
where the term responsible for absorption by the dipole \cite{carminati2006radiative} appears within the brackets. We note  that $\alpha\in\mathds{R}$ is not the appropriate condition for $S$ to be unitary, as could be expected. Instead, one has to include radiative reaction in the discussion, for instance by writing the polarizability as 
\begin{align*}
\alpha=\frac{\alpha_0}{1-\frac{\imath k^3}{6\pi}\alpha_0}~. 
\end{align*}
With this notation, $S$ is unitary if and only if $\alpha_0$ is real. The scattering formalism thus naturally includes the optical theorem as it intrinsically incorporates radiative reaction in the treatment of energy conservation. 

The fact that the operator $D$ for a single dipolar object is directly proportional to the polarizability $\alpha$ shows that the $\mathcal{D}$ operator can be interpreted as a generalization of the polarizability for a more complex system. As a consequence, the unitarity condition for the collective system is a generalization of eq.~(\ref{S_unitary_cond})  
\begin{align}\label{calS_unitary_cond}
I-\calS^\dagger \calS &= -2 \left( \calD  + \calD^\dagger \right) - 4 \calD^\dagger \calD = \mathbf{0} ~ .
\end{align}

\section{Collective absorption} 
Using the collective scattering operator of the ensemble of dipoles, we are now able to study its intrinsic absorption properties. We can in particular define the absorption operator as the deviation of the scattering matrix from unitarity
 which is known to measure the quadratic norm of the noise \cite{genet2003casimir}
\begin{align*}
\mathcal{A} = I- \mathcal{S}^\dagger \mathcal{S} ~.
\end{align*}
The scattering matrix can be rewritten as \cite{fernandez2017unified}
\begin{align*}
\mathcal{S} = U\Sigma V^\dagger = \sum_k \sigma_k u_k v_k^\dagger~,
\end{align*}
with $\Sigma$ a diagonal matrix with real elements $0 \leq \sigma_k \leq 1$, while $U$ and $V$ are the singular vectors that form a basis for \emph{out} and \emph{in} fields, respectively. 
For the absorption operator $\mathcal{A}$ the above decomposition is also valid as 
\begin{align}\label{absorption_svd}
\mathcal{A} = I - V \Sigma^\dagger \Sigma V^\dagger  = \sum_k (1-\sigma_k^2) v_k v_k^\dagger~.
\end{align}
This means that for an incoming mode $v_k$, the fraction $\sigma_k^2$ is re-radiated while a fraction  $\left(1-\sigma_k^2\right)$ is absorbed by the system. 
This decomposition thus gives access to a complete knowledge on the intrinsic absorption properties of the system, by identifying the incoming modes responsible for small or large absorption by the system.

\begin{figure}[ht]
\includegraphics[width=0.45\textwidth]{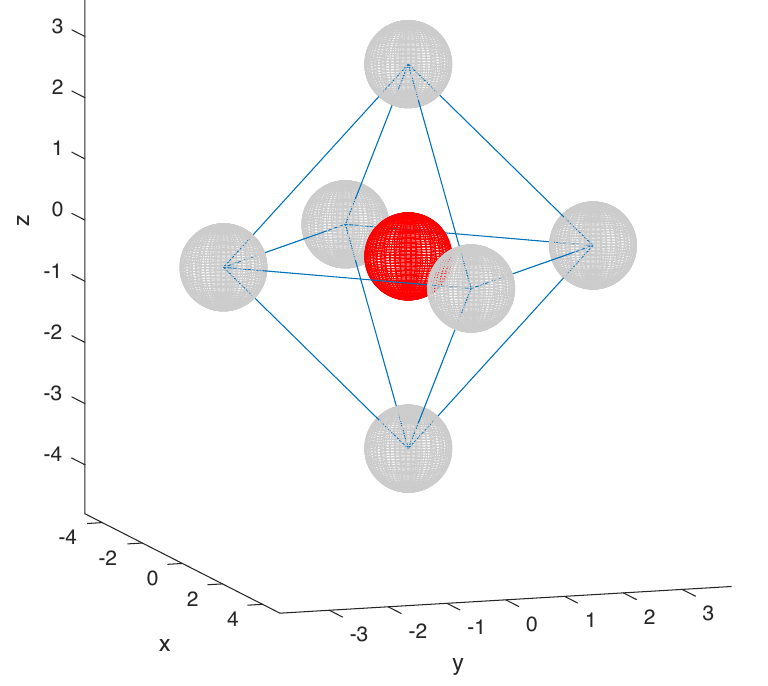} 
\caption{Diagram of the system for $N=6+1$ nanospheres with $a=\lambda/2$ (blue lines are guides to the eye). }
\label{fig:octo}
\end{figure}

We illustrate this property by applying it to a simple system composed of $N$ dielectric nanospheres, where one at the origin is absorbing and the remaining $N-1$ are located at a fixed distance $a$ around it and are non-absorbing. We place them on the vertices of a Platonic solid, which can be a tetrahedron ($N=5$), an octahedron ($N=7$), a cube ($N=9$), an icosahedron ($N=13$) or a dodecahedron ($N=21$), and fix their dielectric function to $\eps=10$, while the central nanosphere features absorption with $\eps_C=\left( \sqrt{10} + 0.1 \imath \right)^2$. The case $N=7$ is sketched on fig.~\ref{fig:octo}. The nanospheres' optical properties are described by a dipolar polarizability obtained from a Clausius-Mossotti model.

We numerically compute the scattering matrix of the collective system with a finite number of spherical modes (up to $\lmax=24$) and extract the singular values and vectors for the absorption operator \eqref{absorption_svd}. We then compare the results to the reference case with the central dipolar absorber only, in order to study the modifications in the absorption by the central dipole due to the non-dissipative outer shell that acts as a collective antenna. We note that the singular value decomposition enables the direct identification of the nature of the modes that yield maximal absorption for the collective system. 

For an isolated dissipative dipole at the origin in the linear optics regime, the absorption is directly proportional to the \emph{incident} electric field amplitude squared at the origin, and therefore only three spherical modes are responsible for a non-zero absorption: the ones associated with coefficients $\Alm$ for $\ell=1$ and $-1\leq m \leq 1$, which form a subspace of $E^\free$ of dimension 3 and are associated with an identical singular value $A_0=1-\sigma_0^2$. For the collective system described above, the absorption is now proportional to the amplitude squared of the \emph{excitation} electric field for the dipole at the origin. As a consequence, the incident modes $\phi_\inc$ responsible for non-zero absorption are the ones that yield a non-zero excitation electric field at the origin. We observe that these incident fields also form a subspace of $E^\free$ of dimension 3, but they can be associated with different singular vectors and singular values $A=1-\sigma^2$, as will be discussed below.

\begin{figure}[ht]
\begin{center}
\includegraphics[width=0.45\textwidth]{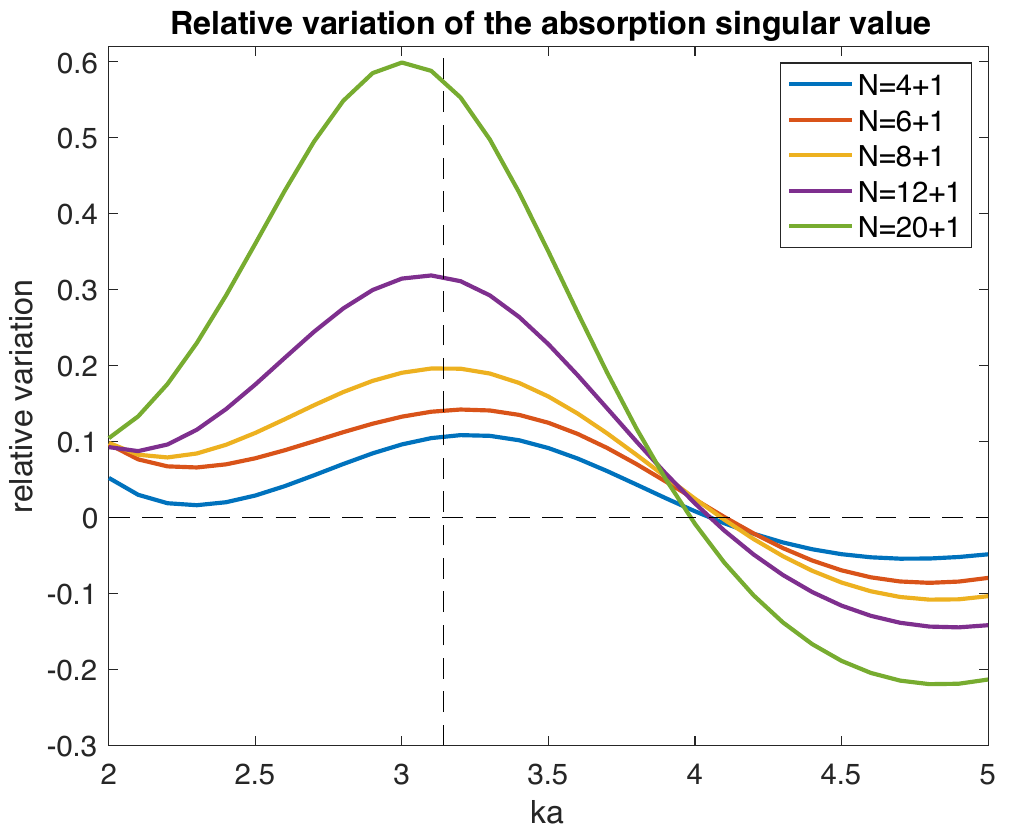} 
\includegraphics[width=0.4\textwidth]{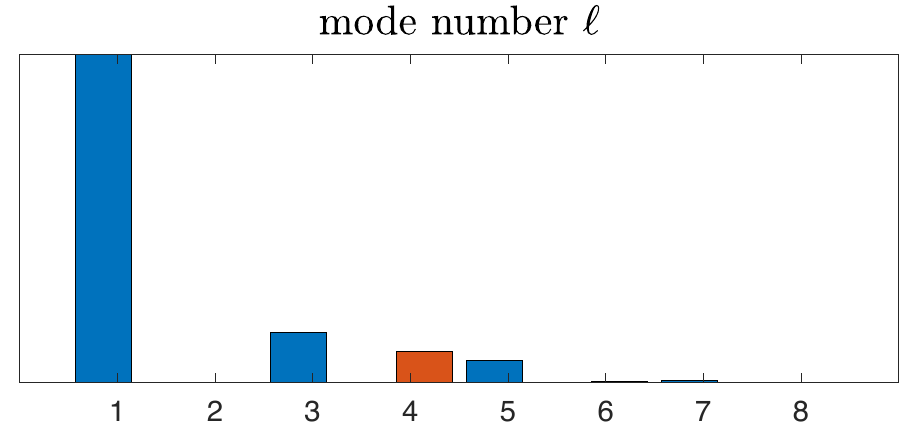} 
\end{center}
\caption{(a) Relative variation of the maximum absorption singular value $A=1-\sigma^2$ as a function of the shell reduced radius $ka$. We consider shells with ($N-1$) nanospheres placed at vertices of Platonic solids with $N-1 \in \{4,6,8,12,20\}$. The nanospheres have a reduced radius $kR=0.8$. The vertical dashed-line corresponds to $a=\lambda/2$. (b) Histogram grouped by values of $\ell$ of the spherical mode coefficients for the singular mode responsible for maximum absorption, in the case of an octahedron with $a=\lambda/2$ (blue bars for $\Alm$ coefficients, red bars for $\Blm$ coefficients).}
\label{fig:abs}
\end{figure}

We present in fig.~\ref{fig:abs}(a) the results of numerical evaluations for the relative change in the maximum singular value for absorption $(A-A_0)/A_0$ due to the presence of the non-dissipative outer shell, as a function of the shell radius $a$. We observe a strong increase in the system absorption for $a\simeq \lambda/2$, which can reach up to $60~\%$ for the dodecahedron. This modification strongly depends on the size of the shell and can result in a decrease of absorption in some cases. In fig.~\ref{fig:abs}(b) we present an histogram of the spherical mode coefficients for the singular mode responsible for maximum absorption for an octahedron with $a=\lambda/2$ which illustrates the appearance of higher multipole components in their composition. These results show that the outer shell acts as an antenna for the central absorber as multiple scattering events enable to amplify the field at its position.

The scattering operator is also a very powerful tool to predict the absorption of the complex system for a series of incoming fields: for a new considered incident field $\phi_\inc$, one only needs to compute the module squared projection onto the three singular modes $(v_{k_1}, v_{k_2}, v_{k_3})$ responsible for absorption and multiply it by the corresponding singular value $A_i$
\begin{align}
\phi_\inc^\dagger \mathcal{A} \phi_\inc &= \phi_\inc^\dagger \left( \sum_k (1-\sigma_k^2) v_k v_k^\dagger \right) \phi_\inc \nonumber \\
&= \sum_k (1-\sigma_k^2)  \left| v_k^\dagger \phi_\inc \right|^2 \nonumber \\
&= \sum_{i=1}^{3} A_i  \left| v_{k_i}^\dagger \phi_\inc \right|^2 ~ .
\end{align}
For instance, for a linearly polarized planar wave (see Appendix \ref{app:planar} for a practical computation of its representation in spherical modes) incoming on a cube shell (N=9) with $a=\lambda/2$, the increase of the absorption singular value $A$ (+$19.7~\%$) observed in fig.~\ref{fig:abs}(a) goes together with an increase of the projection coefficient due to the modification of the singular modes responsible for absorption (+$15.0~\%$) to almost double the increase of effective absorption (+$37.6~\%$) for this specific choice of incident field.

\section{Collective radiative corrections} 
In collective dipolar systems, mutual interactions can lead to dispersive forces as well as energy shifts. In the latter case, the collective shifts can have crucial effect in atomic networks used in a metrological context by limiting the ultimate reachable accuracy. For instance, collective Lamb shifts may result from interactions that shift differently the ground and excited states energy levels of the system.
Here we use the collective scattering formalism to derive the phase shift formula for a collective dipolar system. Applying these results to arrays of small nanoparticles, we show that that collective effect are already quite important when only considering the ground state energy of such an ensemble.

We start from the expression of the collective shift for an ensemble of dipoles as \cite{jaekel1991casimir,plunien1986casimir}
\begin{align}\label{phaseshift_formula}
\Delta(\omega) = \ln \det \mathcal{S}(\omega) - \sum_{i=1}^{N} \ln \det S_i(\omega) ~,
\end{align}
where $\mathcal{S}$ is the scattering matrix of the collective dipolar ensemble and $S_i$ is the scattering matrix for $i$th dipole.  Using the determinant lemma, one deduces from eq.~(\ref{D_dip_result})  
\begin{align*}
\det \mathcal{S} = \frac{ \det \left[ I-X + \begin{pmatrix} \vdots \\  \vspace{-0.1cm} F  T_{i0} \\  \vdots \end{pmatrix}  \begin{pmatrix} \mydots \hspace{-2mm}&, 2(k^3 \alpha_{i'}) T_{0i'} Q ,  & \hspace{-2mm} \mydots \end{pmatrix}  \right]}{\det \left[ I-X \right]} 
\end{align*}
where the matrix appearing in the numerator has diagonal blocks $\left( 1+\imath k^3 \alpha_i /(3\pi) \right) I_3$, $I_3$ being the identity on $\mathds{R}^3$, and non-diagonal blocks $(k^3 \alpha_{i'}) \left[ 2 F T_{i0} T_{0i'} Q - \mathbf{G}_0(\br_i,\br_{i'},\omega)/k \right]$. Writing explicitely the translation operators for $\ell=1$ one shows that 
\begin{align*}
F T_{i0} T_{0i'} Q = \imath \Im \left[ \mathbf{G}_0(\br_i,\br_{i'},\omega)/k\right] ,
\end{align*}
 which is the quantity appearing in the cross density of states (CDOS) \cite{caze2013spatial}. It follows that 
\begin{align}
&\mathbf{G}_0(\br_i,\br_{i'},\omega)/k - 2 F T_{i0} T_{0i'} Q=\mathbf{G}^*_0(\br_i,\br_{i'},\omega)/k ~,
\notag\\
\label{result_Delta}
&  \Delta(\omega) = \ln \det  [I-Y(\omega)] - \ln \det [I-X(\omega)]~,
\end{align}
where $Y$ is a modified structure matrix, similar to $X$ but with non-diagonal blocks $\left[\frac{k^3 \alpha_{i'}}{1+\imath k^3 \alpha_{i'}/(3\pi)} \frac{\mathds{G}_0^*(\br_{i'},\br_j,\omega)}{k}\right]_{i\neq i'}$ instead. In the particular case of non-dissipative dipolar objects, the left fraction in the latter expression is equal to $k^3 \alpha_{i'}^*$, which yields $Y=X^*$ and 
\begin{align}\label{result_Delta_unitary}
  \Delta(\omega)=-2 \imath \Im \left[ \ln \det (I-X(\omega)) \right] ~ .
\end{align}

In such systems, the collective radiative corrections are hence entirely determined by the $3N$ by $3N$ structure matrix $X(\omega)$. Moreover, when looking at frequency-integrated quantity one can use imaginary frequencies $\omega=\imath \xi$ to express for instance dispersive energies as
\begin{align}\label{result_E_intxi}
E=\hbar \int_0^\infty \frac{d \omega}{2\pi} \frac{\imath \Delta(\omega)}{2} = \hbar \int_0^\infty \frac{d\xi}{2\pi} \ln \det [I -X(\imath \xi)] ~ .
\end{align}
The latter result shows that the structure matrix $X$ emerges as the main ingredient to the collective energy shifts resulting from interactions in a dipolar ensemble. 

We now illustrate these results by studying the collective radiative correction for the ground state of an array of $N$ nanoparticles, which in this case corresponds to the Casimir-Polder interaction energy. To highlight collective effects, we compare the collective shift we obtain from (\ref{result_E_intxi}) to the commonly derived pairwise quantity, which is obtained by considering separately all couples of atoms and summing up the radiative corrections obtained for each couple. 
For numerical application we consider separately the cases of 1D, 2D and 3D arrays, all being made of gold nanoparticles with radius $R=20$ nm and optical properties modeled by a plasma model.

\begin{figure}[htbp]
\includegraphics[width=0.45\textwidth]{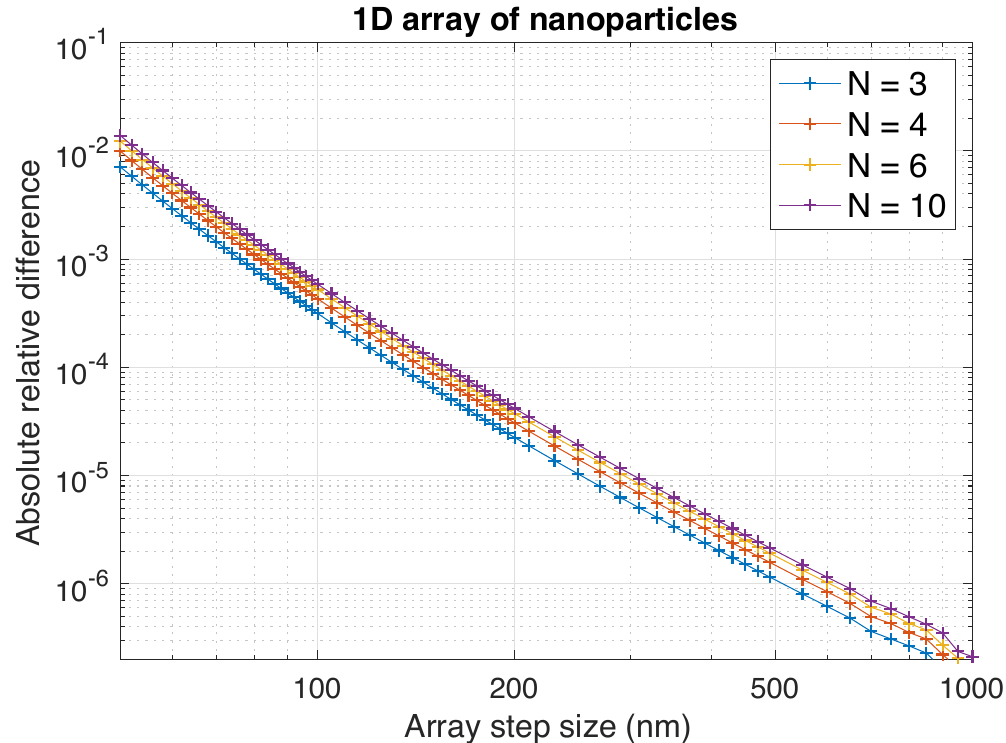} 

\vspace{0.5cm}

\includegraphics[width=0.45\textwidth]{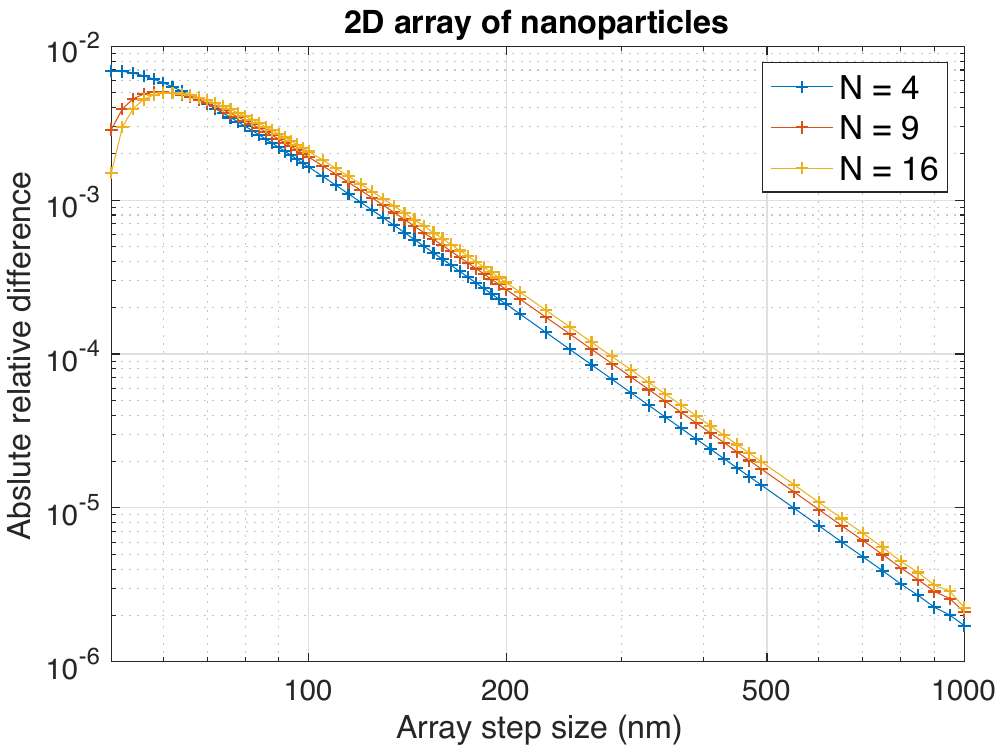} 

\vspace{0.5cm}

\includegraphics[width=0.45\textwidth]{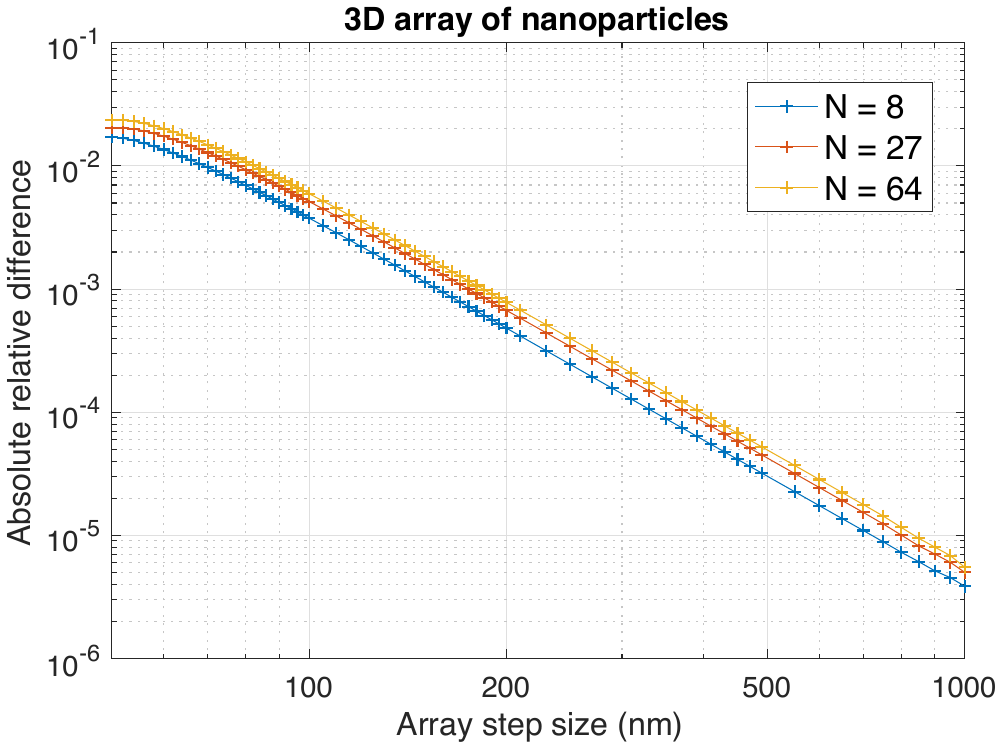} 
\caption{Absolute relative difference between the collective dispersive energy $E$ and the associated pairwise quantity $E_\mathrm{PW}$ as a function of the array step size, for 1D array (a) 2D array (b) and 3D array (c).}
\label{fig:Delta_CP_arrays}
\end{figure}

For collective systems with $N>2$, we observe a deviation from the pairwise quantity, which increases when the array step size is reduced, as shown in fig.~\ref{fig:Delta_CP_arrays}. This deviation is also increased by the number of atoms in the array, as expected, and the convergence between the two results at large spacings is slower for higher-dimensional arrays. In the case of a 3D atomic array, deviations of the order of the percent are obtained for array spacings up to 100 nm. 

These results show that for simple dipolar systems, collective effects are already quite important when studying the radiative correction for ground-state energy. For collective dipolar systems, such as optical atomic clocks, where the collective shift of the transition energy is at play, the latter could be obtained by combining shifts for the ground-state as well as for the considered excited state. As the latter is expected to be of much longer range due to the slow decay of interactions between excited states \cite{donaire2015quasiresonant,donaire2016twoatom,jentschura2017virtual,jentschura2017long}, the presented results show that collective effects could play a major role in the energy shifts for atomic optical clocks, as already foreseen. 

\section*{Conclusions}
We have introduced a theoretical formalism merging the Coupled Dipole Model approach with spherical electromagnetic modes to derive the collective scattering  matrix for an ensemble of dipoles. This enables us to study the intrinsic optical properties of a collective dipolar system, such as its absorption, without specifying a particular illumination condition. The scattering formalism is also well connected to the study of dispersive interaction energy between scatterers, the presented approach thus brings the opportunity to investigate collective effects in such interactions. In particular, the collective scattering formalism presented in this article, combined with a treatment of excited states, could be a way to investigate the collective energy shifts in atomic arrays, a topic of major importance in the development of collective atomic clocks.

\section*{Acknowledgments} 
Thanks are due to T. Ebbesen, C. Genet and R. Gu\'erout for stimulating discussions.


\appendix

\section{Explicit expressions for the used operators and functions}

\subsection{Green tensor in vacuum}
\label{app:green}

The 3D Green tensor in vacuum between locations $\br$ (source) and $\br'$ (target) is
\begin{align}\label{app_green}
\frac{\mathbf{G}_0(\br',\br,\omega)}{k} = \frac{e^{\imath kr}}{4\pi kr} &\left[\frac{(kr)^2+\imath kr -1}{(kr)^2} \Id \right. \nonumber \\
&\left. - \frac{(kr)^2+3 \imath kr -3}{(kr)^2} \bu_r \otimes \bu_r \right]
\end{align}
where $k=\omega/c$, $r=\| \br' - \br\|$, $\bu_r = (\br' - \br)/r$ and $\bu_r \otimes \bu_r$ is the tensor product of vector $\bu_r$ with itself such that for any $\bu$, $\left( \bu_r \otimes \bu_r \right) \cdot \bu = \bu_r \left( \bu_r \cdot \bu \right)$.

\subsection{Vector spherical modes}
\label{app:spherical}

Any harmonic electromagnetic field in vacuum can be decomposed into vector spherical modes as
\begin{align}\label{def_EH_app}
\bE(\br) = &\frac{P_0}{\sqrt{\eps_0}} \sum_{\ell=1}^{\infty} \sum_{m=-\ell}^{\ell} A_{\ell,m}^{(1)} \bN_{\ell,m}^{(1)}(\br) + A_{\ell,m}^{(2)} \bN_{\ell,m}^{(2)}(\br) \nonumber \\
 & ~  ~  ~  ~ + \imath B_{\ell,m}^{(1)} \bMM_{\ell,m}^{(1)}(\br) + \imath B_{\ell,m}^{(2)} \bMM_{\ell,m}^{(2)} (\br)\nonumber \\
\bH(\br) = &\frac{P_0}{\sqrt{\mu_0}} \sum_{\ell=1}^{\infty} \sum_{m=-\ell}^{\ell} B_{\ell,m}^{(1)} \bN_{\ell,m}^{(1)}(\br) + B_{\ell,m}^{(2)} \bN_{\ell,m}^{(2)}(\br) \nonumber \\
& ~ ~  ~  ~ - \imath A_{\ell,m}^{(1)} \bMM_{\ell,m}^{(1)}(\br) - \imath A_{\ell,m}^{(2)} \bMM_{\ell,m}^{(2)}(\br)
\end{align}
where the prefactor $P_0$ is a normalizing amplitude, $\eps_0$ and $\mu_0$ are the vacuum electric permittivity and magnetic permeability, and the electromagnetic spherical modes are, in spherical coordinates ($r, \theta, \phi$):
\begin{align*}
\bMM^{(q)}_{\ell,m} (\br) &= \frac{1}{kr\sqrt{\ell(\ell+1)}} \begin{pmatrix} 0 \\ \\ \displaystyle -\frac{m}{\sin\theta} \xi^{(q)}_\ell(kr) \Ylm (\theta,\phi) \\ \\ \displaystyle - \imath \xi^{(q)}_\ell(kr) \partial_\theta \Ylm (\theta,\phi) \end{pmatrix} 
\end{align*}
\begin{align*}
\bN^{(q)}_{\ell,m} (\br) &=  \frac{1}{kr\sqrt{\ell(\ell+1)}}  \begin{pmatrix} \displaystyle \imath \frac{\ell (\ell+1)}{kr}  \xi^{(q)}_\ell(kr) \Ylm (\theta,\phi) \\ \\ \displaystyle \imath \partial_{(kr)} \xi^{(q)}_\ell(kr) \partial_\theta \Ylm (\theta,\phi) \\ \\ \displaystyle -\frac{m}{\sin\theta}  \partial_{(kr)} \xi^{(q)}_\ell(kr) \Ylm (\theta,\phi)  \end{pmatrix} ~ ,
\end{align*}
with the indices $\ell\geq1$ and $-\ell \geq m \geq \ell$, where $\xi_\ell^{(q)}$ are the Riccati-Hankel functions with $q=1$ for modes going outwards and $q=2$ for modes coming inwards. In the context of free fields, one can alternatively considers the modes $\bN^{(j)}_{\ell,m}$ where the Riccati-Hankel functions $\xi_\ell^{(q)}$ are replaced by the Riccati-Bessel functions $\psi_\ell$. 

The spherical harmonics $\Ylm$ are defined as
\begin{align}\label{def_Ylm}
\Ylm (\theta,\phi) = \sqrt{\frac{2\ell+1}{4\pi} \frac{(\ell-m)!}{(\ell+m)!}} ~ e^{\imath m \phi} ~ \Plm(\cos\theta)
\end{align}
where $\Plm$ are the associated Legendre polynomials and the normalizing factor is chosen such that the functions obey the orthogonality relation
\begin{align}\label{ortho_Ylm}
\int_\Omega Y_{\ell_1,m_1} (\theta,\phi) Y_{\ell_2,m_2}^* (\theta,\phi) \sin \theta \d \theta \d \phi = \delta_{\ell_1,\ell_2} \delta_{m_1,m_2} 
\end{align}
where $\Omega$ is the full solid angle with $\theta\in[0 , \pi]$ and $\phi \in [0, 2\pi]$. This enables to express quite simply the coefficients of the decomposition of any electromagnetic fields in $E_\free$ or $E_\exit$. Indeed, when writing the radial component of the electric field on a sphere of arbitrary radius $a$ one gets
\begin{align*}
\frac{\bE(a,\theta,\phi) \cdot \bu_r}{P_0/\sqrt{\eps_0}} = \sum_{\ell=1}^{\infty} \sum_{m=-\ell}^{\ell} & \frac{\imath \sqrt{\ell(\ell+1)}}{(ka)^2} \Ylm (\theta,\phi)  \\ 
& \times \left[ \Alm^{(1)} \xi_\ell^{(1)} (ka) + \Alm^{(2)} \xi_\ell^{(2)}(ka)\right] 
\end{align*}
and a similar result is obtained for the magnetic field normalized by $P_0/\sqrt{\mu_0}$ with the coefficients $\Blm$. If one now assumes that this vector field is an outgoing field in $E_\exit$, one has $ \Alm^{(2)}= \Blm^{(2)} = 0, \forall \ell,m$ and the coefficients can be determined directly using the orthogonality relation (\ref{ortho_Ylm}):
\begin{align}
\Alm^{(1)} = &\frac{(ka)^2}{\imath \sqrt{\ell(\ell+1)} \xi^{(1)}_l(ka)} \nonumber \\
& \times \int_\Omega~\frac{\bE(a,\theta,\phi)\cdot \bu_r}{P_0/\sqrt{\eps_0}} \Ylm^*(\theta,\phi)  \sin\theta \d \theta \d \phi  \label{decompo_Alm_out} \\
\Blm^{(1)} = &\frac{(ka)^2}{\imath \sqrt{\ell(\ell+1)} \xi^{(1)}_l(ka)} \nonumber \\
& \times \int_\Omega~\frac{\bH(a,\theta,\phi)\cdot \bu_r}{P_0/\sqrt{\mu_0}} \Ylm^*(\theta,\phi)  \sin\theta \d \theta \d \phi \label{decompo_Blm_out} ~ .
\end{align}
Similarly, the coefficients $\Alm^{(j)}, \Blm^{(j)}$ of a free field can be obtained directly replacing $\xi^{(1)}_l (ka)$ by $\psi_l (ka)$ in the denominators of the latter expressions. For more general vector fields the decomposition is more demanding and involves either non-radial components of the fields or derivatives of their radial components in order to determine separately the coefficients $\Alm^{(1)},\Alm^{(2)}$.

\subsection{Decomposition of an incident planar wave into spherical modes} 
\label{app:planar}

In this section we consider an incident planar field that propagates along the $z$-direction with linear polarization:
\begin{align*}
&\bE (\br) = \frac{P_0}{\sqrt{\eps_0}} e^{\imath k z} \left( \begin{array}{c} 1 \\ 0 \\ 0 \end{array} \right) 
&\bH (\br) = \frac{P_0}{\sqrt{\mu_0}} e^{\imath k z} \left( \begin{array}{c} 0 \\ 1 \\ 0 \end{array} \right) 
\end{align*}
which is a free field, and aim at representing this electromagnetic field in spherical modes. Using (\ref{decompo_Alm_out}, \ref{decompo_Blm_out}) for a free field, we obtain the coefficients from the radial components as separate integrals:
\begin{align*}
\Alm^{(j)} &= 
 \frac{(ka)^2}{\imath \sqrt{\ell(\ell+1)} \psi_l(ka)} \sqrt{\frac{2\ell+1}{4\pi}} \sqrt{\frac{(\ell-m)!}{(\ell+m)!}} \\
& ~ ~ ~ \times \left[ \int_0^{2\pi}  e^{-\imath m \phi} \cos \phi ~ \d \phi \right] \\
& ~ ~ ~ ~ ~ \times \left[ \int_0^\pi \sin^2\theta  e^{\imath k a \cos \theta} \Plm(\cos\theta) \d \theta \right] \\
\Blm^{(j)} &= 
\frac{(ka)^2}{\imath \sqrt{\ell(\ell+1)} \psi_l(ka)} \sqrt{\frac{2\ell+1}{4\pi}} \sqrt{\frac{(\ell-m)!}{(\ell+m)!}} \\
& ~ ~ ~ \times \left[ \int_0^{2\pi} e^{-\imath m \phi} \sin \phi ~\d \phi \right] \\
& ~ ~ ~ ~ ~ \times \left[ \int_0^\pi \sin^2\theta e^{\imath k a \cos \theta} \Plm(\cos\theta) \d \theta  \right] ~ .
\end{align*} 
The integrals over $\phi$ only give nonzero results for $m=\pm1$:
\begin{align}\label{result_planarwave}
\int_0^{2\pi} e^{-\imath m \phi} \cos \phi ~\d \phi &= \left\{ \begin{array}{rcl} \pi & \mathrm{for} & m=\pm 1 \\  0 & \mathrm{for} & |m|\neq 1 \end{array} \right. \nonumber  \\
\int_0^{2\pi} e^{-\imath m \phi} \sin \phi  ~\d \phi &= \left\{ \begin{array}{rcl} -\imath m \pi & \mathrm{for} & m = \pm 1 \\  0 & \mathrm{for} & |m|\neq 1 \end{array} \right. 
\end{align}
and the integrals over $\theta$ only need to be evaluated for $m=\pm1$. For $m=1$, recalling that $P_\ell^1 (\cos\theta) = \partial_\theta \left[ P_\ell (\cos \theta)\right]$ and using properties of Legendre polynomials, the integral over $\theta$ can be computed with integration by parts, yielding
\begin{align*}
I_{(m=1)} 
&= \frac{\imath \ell (\ell+1)}{ka} \int_0^\pi e^{\imath k a \cos \theta } P_\ell(\cos\theta) \sin\theta \d \theta \\
&= 2(\imath)^{\ell+1}\ell(\ell+1)  \frac{j_\ell(ka)}{ka}
\end{align*}
where we have used Poisson's integral for the last step. The $m=-1$ case can be derived directly from this result using $P_\ell^{-1}(\cos\theta)=-\frac{1}{\ell(\ell+1)} P_\ell^1(\cos\theta)$ and gives
\begin{align*}
I_{(m=-1)} &= -2(\imath)^{\ell+1}  \frac{j_\ell(ka)}{ka} ~ .
\end{align*}
One can finally gather the results for the two integrals to get the final expressions for the multipolar coefficients for this linearly polarized planar incident field:
\begin{align*}
\Alm^{(j)} &=  \left\{ \begin{array}{rcl} m (\imath)^\ell \sqrt{\pi(2\ell+1)} & \mathrm{for} & m=\pm 1 \\  0 & \mathrm{for} & |m|\neq 1 \end{array} \right. \\
\Blm^{(j)} &=  \left\{ \begin{array}{rcl} (\imath)^{\ell-1} \sqrt{\pi(2\ell+1) } & \mathrm{for} & m=\pm 1  \\  0 & \mathrm{for} & |m|\neq 1 \end{array} \right. ~ . \\
\end{align*}

~

\subsection{Translation operators in spherical coordinates}
\label{app:Tmatrix}

In this section we give explicit expressions for the translations operators $T_{i0}, T_{0i}$ following the works of \cite{wittmann1988spherical} with minor changes in the notations to be consistent with the present paper and with a more practical implementation of the recurrent definition for the Clebsch-Gordan coefficients. We start by considering an electromagnetic field as defined in Eq.~(\ref{def_EH_app}) that is either a free field $\phi \in E^\free$ or an outgoing field $\phi \in E^\exit$
\begin{align*}
\bE(\br) &= \frac{P_0}{\sqrt{\eps_0}} \sum_{\ell,m} A_{\ell,m} \bN_{\ell,m}(\br) + \imath B_{\ell,m} \bMM_{\ell,m}(\br) 
\end{align*}
where $\bN_{\ell,m}$ is $\bN^{(j)}_{\ell,m}$ for a free field and $\bN^{(1)}_{\ell,m}$ for an outgoing field. Once translated by a vector $\brho$, it can also be decomposed into spherical modes:
\begin{align*}
\bE(\br-\brho) &=\frac{P_0}{\sqrt{\eps_0}} \sum_{\ell,m} A^{'}_{\ell,m}  \bN_{\ell,m}(\br) + \imath B^{'}_{\ell,m} \bMM_{\ell,m}(\br)
\end{align*}
which can also a be a free field or an outgoing field, depending on the kind of translation we consider. In any case the new coefficients can be obtained from the old ones by a linear relation written with compact notations:
\begin{align*}
\begin{pmatrix} A_{\ell,m}^{'} \\ B_{\ell,m}^{'} \end{pmatrix}
= \begin{pmatrix} T^B  & \imath T^C  \\ - \imath T^C  & T^B  \end{pmatrix}
 \begin{pmatrix}A_{\ell,m} \\ B_{\ell,m}\end{pmatrix} 
\end{align*}
where $T^B$ and $T^C$ are square matrices of size $N_\sph/2$ by $N_\sph/2$. When we consider a translation of a free field into another free field, as realized by the translation operators $T_{i0}$ that are used to translate the global incident field $\phi_\inc$, we will call these blocks $T^B_\free, T^C_\free$. When we instead consider the translation of an outgoing field into another outgoing field, as realized by the operators $T_{0i}$ that transforms a field $\phi_{\sca,i}$ scattered by a sphere into its contribution to the total scattered field $\phi_\sca$ by the global system, we will call these blocks as $T^B_\exit, T^C_\exit$. These block operators happen to be equal, we will thus refer to them as $T^B$ and $T^C$ only in the following. The explicit expression of these blocks' elements are:
\begin{widetext}
\begin{align*}
&T^B_{\ell_1, m_1, \ell_2, m_2} (\brho) = (-1)^{m_1} 4 \pi (\imath)^{\ell_2-\ell_1} \sum_{\alpha} \frac{\ell_1(\ell_1+1) + \ell_2(\ell_2+1) - \alpha(\alpha+1)}{2\sqrt{\ell_1(\ell_1+1)\ell_2(\ell_2+1)}}  \imath^{\alpha} a(\alpha, \beta | \ell_1,  -m_1,\ell_2,m_2) u_{\alpha,\beta} (\brho) \nonumber \\
&T^C_{\ell_1, m_1, \ell_2, m_2} (\brho) = \frac{-\imath k \brho}{\sqrt{\ell_1(\ell_1+1)\ell_2(\ell_2+1)}} \cdot 
\begin{pmatrix}
\left[ \lambda_{\ell_2,m_2}^{(+)} T^A_{\ell_1,m_1,\ell_2,m_2+1} (\brho) + \lambda_{\ell_2,m_2}^{(-)} T^A_{\ell_1,m_1,\ell_2,m_2-1} (\brho)\right] / 2 \\
\left[ \lambda_{\ell_2,m_2}^{(+)} T^A_{\ell_1,m_1,\ell_2,m_2+1} (\brho) - \lambda_{\ell_2,m_2}^{(-)} T^A_{\ell_1,m_1,\ell_2,m_2-1} (\brho)\right] / (2\imath) \\
m_2 T^A_{\ell_1,m_1,\ell_2,m_2} (\brho)  
\end{pmatrix}
\end{align*}
%
where the sum over $\alpha$ is carried with $|\ell_1-\ell_2| \leq \alpha \leq \ell_1+\ell_2$ and $\beta=m_2-m_1$. Here we have defined $u_{\ell,m} (\br) = j_\ell (k r) Y_{\ell,m} (\theta,\phi)$, with $j_\ell (x)=\psi_\ell(x)/x$ the spherical Bessel functions, the prefactors $\lambda^{(\pm)}_{\ell,m} = \sqrt{(\ell \mp m)(\ell \pm m+1)}$ and the blocks $T^A$:
%
\begin{align*}
T^A_{\ell_1,m_1,\ell_2,m_2} (\brho) = (-1)^{m_1} 4\pi (\imath)^{\ell_2-\ell_1} \sum_{\alpha} \imath^{\alpha} a(\alpha, \beta | \ell_1,  -m_1,\ell_2,m_2) u_{\alpha,\beta} (\brho)
\end{align*}
where the sum over $\alpha$ is again carried with $|\ell_1-\ell_2| \leq \alpha \leq \ell_1+\ell_2$ and $\beta=m_2-m_1$. These coefficients also use the coefficients $a$:
\begin{align*}
a(\alpha, \beta | \ell_1,  m_1,\ell_2,m_2) = \sqrt{\frac{(2\ell_1+1)(2\ell_2+1)}{4\pi (2\alpha+1)}} \left< \alpha, \beta | \ell_1, m_1,\ell_2,m_2\right> \left< \alpha, 0 | \ell_1, 0 , \ell_2,  0 \right> ~ .
\end{align*}
%
The latter expression in turn involves the Clebsch-Gordan coefficients $\left< \alpha, \beta | \ell_1, m_1,\ell_2,m_2\right>$ with $\beta=m_2+m_1$ that can be defined by downward recurrence on $\alpha$ from $(\ell_1+\ell_2)$ down to $0$ with
%
\begin{align*}
\left< \alpha, \beta | \ell_1, m_1,\ell_2,m_2\right> &= 0 ~ ~ \mathrm{when} ~ ~ \alpha>\ell_1+\ell_2 \\
\left< (\ell_1+\ell_2), \beta | \ell_1, m_1,\ell_2,m_2\right> &=\sqrt{\frac{(\ell_1+\ell_2+m_1+m_2)!(\ell_1+\ell_2-m_1-m_2)!}{(\ell_1+m_1)!(\ell_1-m_1)!(\ell_2+m_2)!(\ell_2-m_2)!}} \sqrt{\frac{(2\ell_1)!(2\ell_2)!}{(2\ell_1+2\ell_2)!}} \\
\left< \alpha, \beta | \ell_1, m_1,\ell_2,m_2\right> &= \frac{\zeta_\alpha}{\sqrt{\xi_\alpha}} \left< \alpha+1, \beta | \ell_1, m_1,\ell_2,m_2\right> - \sqrt{\frac{\xi_{\alpha+1}}{\xi_\alpha}} \left< \alpha+2, \beta | \ell_1, m_1,\ell_2,m_2\right>
\end{align*}
where
\begin{align*}
\zeta_\alpha &= m_1-m_2 - \beta\frac{\ell_1(\ell_1+1)-\ell_2(\ell_2+1)}{(\alpha+1)(\alpha+2)} \\
\xi_\alpha &=\frac{\left[ (\alpha+1)^2 - \beta^2 \right]\left[ (\alpha+1)^2 - (\ell_2-\ell_1)^2 \right]\left[(\ell_1+ \ell_2+1)^2 - (\alpha+1)^2  \right]}{(\alpha+1)^2 \left[ 4(\alpha+1)^2 - 1\right]} ~ .
\end{align*}
\end{widetext}
%


%

Finally, we get the translation operators in spherical modes:
\begin{align*}
T_{i 0}&=  \begin{pmatrix} T^B(\br_i)  & \imath T^C(\br_i)  \\ - \imath T^C(\br_i)  & T^B(\br_i)  \end{pmatrix}\\
T_{0 i} &=  \begin{pmatrix} T^B(-\br_i)  & \imath T^C(-\br_i) \\ - \imath T^C (-\br_i) & T^B (-\br_i) \end{pmatrix} ~ .
\end{align*}
%

\subsection{$F$ operator: evaluation of an electromagnetic field from its spherical decomposition}
\label{app:Fmatrix}

The operator $F$ is a $3$-by-$N_\sph$ matrix that evaluates the electric field $\bE$ of a free field at the origin from its spherical mode decomposition $\phi$. Before evaluating the spherical modes $\bN_{\ell,m}^{(j)}, \bMM_{\ell,m}^{(j)}$ at the origin, we first remark that the Riccati-Bessel functions $\psi_\ell(x)$ and their derivatives have the small-$x$ expansions:
\begin{align*}
\psi_\ell(x) \sim \frac{\sqrt{\pi}}{\Gamma(\ell+3/2)} \left( \frac{x}{2} \right)^{\ell+1} ~ ;  ~
\psi^{'}_\ell(x) \sim \frac{(\ell+1) \sqrt{\pi}}{2 \Gamma(\ell+3/2)} \left( \frac{x}{2}\right)^\ell  ~ .
\end{align*} 
As a consequence, the components of $\bMM_{\ell,m}^{(j)}$, which scale as $\psi_\ell(x)/x$ with $x=kr$, all vanish at the origin as $\ell\geq1$. The components of $\bN_{\ell,m}^{(j)}$, which scale as $\psi_\ell(x)/x^2$ and $\psi^{'}_\ell(x)/x$, have non-zero limits only for $\ell=1$, for which one has
\begin{align*}
&\psi_1(x) \sim \frac{x^2}{3}
&\psi^{'}_1(x) \sim \frac{2x}{3} ~ .
\end{align*} 
We then need the explicit expressions for the spherical harmonics $\Ylm$ and their derivatives $\partial_\theta \Ylm$ with $\ell=1$, which are
\begin{align*}
&Y_{1,-1} (\theta,\phi) = \sqrt{\frac{3}{8\pi}} \sin\theta e^{-\imath \phi} \\
& Y_{1,0} (\theta,\phi) =  \sqrt{\frac{3}{4\pi}} \cos\theta  \\
&Y_{1,1}(\theta,\phi) = - \sqrt{\frac{3}{8\pi}} \sin\theta e^{\imath \phi} \\
& \partial_\theta Y_{1, -1}(\theta,\phi) = \sqrt{\frac{3}{8\pi}} \cos\theta  e^{-\imath \phi} \\
& \partial_\theta Y_{1,0} (\theta,\phi) =  - \sqrt{\frac{3}{4\pi}} \sin\theta  \\
&\partial_\theta Y_{1,1} (\theta,\phi) = - \sqrt{\frac{3}{8\pi}} \cos\theta e^{\imath \phi}
\end{align*}
from which we find that the spherical modes $\bN_{1,m}^{(j)}$ evaluated at the origin give
\begin{align*}
&\bN_{1,-1}^{(j)}(\mathbf{0}) = \frac{1}{\sqrt{12\pi}} \begin{pmatrix} \imath \\ 1 \\ 0 \end{pmatrix}
&\bN_{1,0}^{(j)}(\mathbf{0}) = \frac{1}{\sqrt{6\pi}} \begin{pmatrix} 0 \\ 0 \\ \imath \end{pmatrix} \\
&\bN_{1,-1}^{(j)}(\mathbf{0}) = \frac{1}{\sqrt{12\pi}} \begin{pmatrix} -\imath \\ 1 \\ 0 \end{pmatrix} 
&
\end{align*}
where the vectors are written in Cartesian coordinates.

Finally, the evaluation of a free field $\phi$ at the origin can be realized by a multiplication by the matrix $F$ as 
\begin{align}\label{Fmatrix_def}
&\frac{\bE(\mathbf{0})}{P_0/\sqrt{\eps_0}} = \underbrace{\frac{1}{\sqrt{12\pi}} \begin{pmatrix} 
 \imath & 0 & -\imath & 0 & \hdots \\
 1 & 0 & 1 & 0 & \hdots \\
 0 & \sqrt{2} \imath & 0 & 0 & \hdots  \end{pmatrix}}_{F}
\phi
\end{align}
where the only non-zero coefficients correspond to $\Alm^{(j)}$ with $\ell=1$ and $m\in \{-1,0,1\}$. Using (\ref{def_EH_app}), the same expression can obtained for the magnetic field with a matrix $F_{(H)}$ where the three same columns are corresponding the $\Blm^{(j)}$ coefficients with $\ell=1$ and $m\in \{-1,0,1\}$. Let us finally remark that $\sqrt{6\pi} F$ is pseudo-unitary, in the sense that
\begin{align*}
F F^\dagger = \frac{I}{6 \pi} 
\end{align*}
where the identity acts on $\mathds{R}^3$, while $F^\dagger F=I_{(A,\ell=1)}/(6\pi)$ with $I_{(A,\ell=1)}$ the identity operator for coefficients $A_{\ell,m}$ with $\ell=1, m\in \{-1,0, 1 \}$ and zero elsewhere.

\subsection{$Q$ operator: multipolar expansion of dipolar emission}
\label{app:Qmatrix}

The operator $Q$ is a $N_\sph$-by-3 matrix that evaluates the coefficients $\Alm^{(1)}$, $\Blm^{(1)}$ of the field $\phi_\sca^{0}$ generated by a dipole at the origin with orientation $\bu \in \mathds{R}^3$ by
\begin{align}\label{def_Q}
\phi_\sca^{0} = Q \cdot \bu ~ .
\end{align}
 We first consider a dipole whose moment is aligned in the $z$-direction with $\bu = \bu_z$. The electric field radiated by the dipole is expressed thanks to the vacuum Green tensor (\ref{app_green}) as
\begin{align*}
\bE^{[z]}(\br) &= \mu_0 \omega^2 \mathbf{G}_0 (\br,\mathbf{0}, \omega) \cdot p_0 \bu_z \\
&= \frac{P_0}{\sqrt{\eps_0}} \frac{\mathbf{G}_0(\br,\mathbf{0}, \omega) \cdot  \bu_z}{k} ~  ~ \mathrm{with} ~ ~ P_0 = \frac{k^3 p_0}{\sqrt{\eps_0}}
\end{align*}
Then
\begin{align*}
\frac{\bE^{[z]}(\br)}{P_0/\sqrt{\eps_0}} = \frac{e^{\imath kr}}{4\pi kr} &\left[\frac{(kr)^2+\imath kr -1}{(kr)^2} \bu_z \right. \\
 & \left. - \frac{(kr)^2+3 \imath kr -3}{(kr)^2} \cos \theta~ \bu_r  \right]
\end{align*}
from which one gets the components of the electric field in spherical coordinates as
\begin{align}
\frac{\bE^{[z]}(\br) \cdot \bu_r}{P_0/\sqrt{\eps_0}} &= \frac{e^{\imath kr}}{4\pi k r} \cos \theta \left[ f_1(kr) - f_2 (kr) \right]  \label{Eradial_f1f2}\\
\frac{\bE^{[z]}(\br) \cdot \bu_\theta}{P_0/\sqrt{\eps_0}} &= - \frac{e^{\imath kr}}{4\pi k r} \sin \theta  f_1(kr) \\
\frac{\bE^{[z]}(\br) \cdot \bu_\phi}{P_0/\sqrt{\eps_0}} &= 0
\end{align} 
where we have used the fonctions $f_1(x)=(x^2+\imath x-1)/x^2$ and $f_2(x)=(x^2+3\imath x-3)/x^2$. The magnetic field $\bH^{[z]}$ is related to  the electric field by $\nabla \times \bE^{[z]} = \imath \mu_0 \omega \bH^{[z]}$, which gives a vanishing radial component:
\begin{align*}
\frac{\bH^{[z]}(\br) \cdot \bu_r}{P_0/\sqrt{\mu_0}} = \frac{1}{\imath k r\sin\theta} \left[ \frac{\partial}{\partial \theta} \frac{\sin\theta E^{[z]}_\phi}{P_0/\sqrt{\eps_0}} - \frac{\partial}{\partial \phi} \frac{E^{[z]}_\theta}{P_0/\sqrt{\eps_0}}\right]  = 0 ~ .
\end{align*}
Unlike the electric field, the magnetic field is then purely transverse, even for small distances to the emitter where the near-field has an important role. Furthermore, using (\ref{decompo_Blm_out}) this shows that the the coefficients $\Blm^{(1), [z]}$ for spherical magnetic modes will all be zero for the radiated field. This property of transversality is independent on the direction of the dipole momentum and will therefore be obtained for any orientation of the emissive dipole. 

Coming back to the electric field components, the radial component (\ref{Eradial_f1f2}) can be used to determine the coefficients $\Alm^{(1), [z]}$ with formula (\ref{decompo_Alm_out}) as the field is an outgoing mode. By substitution one gets
\begin{align*}
\Alm^{(1), [z]} = &\frac{ e^{\imath ka}}{2\imath \pi \sqrt{\ell(\ell+1)} \xi_\ell^{(1)}(ka)} \frac{-\imath ka +1}{ka} \\
& \times \int_\Omega \cos\theta \sin \theta ~\Ylm^{*}(\theta,\phi) \d \theta \d \phi 
\end{align*}
and we compute the later integral over the solid angle by separating the two integrations as
\begin{align*}
I_{\ell,m}^{[z]} = &\sqrt{\frac{2\ell+1}{4\pi}} \sqrt{\frac{(\ell-m)!}{(\ell+m)!}} \int_0^\pi \cos\theta \sin \theta P_\ell^m(\cos \theta) ~\d \theta \\
&\times  \int_0^{2\pi}  e^{-\imath m \phi} ~\d \phi ~ .
\end{align*}
The integral over $\phi$ is zero except for $m=0$, for which it yields the value $2\pi$. The integral over $\theta$ can then be derived using the orthogonality relation of associated Legendre fonctions
\begin{align}\label{ortho_Plm}
\int_0^\pi P_\ell^m(\cos\theta) P_{\ell'}^m(\cos\theta) \sin\theta\d \theta = \frac{2 (\ell+m)!}{(2\ell+1) (\ell-m)!} \delta_{\ell,\ell'} ~ .
\end{align}
For $m=0$ and $\ell'=1$, one has $P_1^0(\cos\theta) = \cos\theta$ and this relation leads to
\begin{align*}
\int_0^\pi \cos \theta \sin\theta P_\ell^0 (\cos\theta) \d \theta = \frac{2}{2\ell+1} \delta_{1,\ell}
\end{align*}
which is $2/3$ for $\ell=1$ and $0$ when $\ell > 1$. It follows that the quantity $I_{\ell,m}^{[z]}$ is $\sqrt{4\pi/3}$ for $\ell=1, m=0$ and zero for other values of the indices. For $\ell=1$ the Riccati-Hankel function is $\xi_1^{(1)}(ka) = -\imath (-\imath ka+1)e^{\imath ka}/(ka)$ and as a consequence the coefficient $A_{10}^{(1), [z]}$ does not depend on the chosen radius $a$. Finally the only non-zero coefficient of the spherical electric modes for a vector field radiated by a dipole whose momentum is along the $z$-direction writes
\begin{align}\label{Alm_pz}
A_{1,0}^{(1),[z]} = \frac{I_{1,0}^{[z]}}{2\sqrt{2} \pi} = \frac{1}{\sqrt{6 \pi}}
\end{align} 
and vanishes for all other values of the indices $(\ell,m)$.

We now consider the cases where the dipole is still located at the origin, but its momentum can have components along the $x$ and $y$ directions. 
As the transversality of the magnetic field is a general property for the radiation of an electric dipole, the coefficients $\Blm^{(1), [x]}$ and $\Blm^{(1), [y]}$ are zero. We therefore only consider the electric field in order to derive the $\Alm^{(1), [x]}$ and $\Alm^{(1),[y]}$ coefficients.

First, by applying the directions $\bu_x$ and $\bu_y$ on the right of the Green tensor, one gets
\begin{align*}
\frac{\mathbf{G}_0(\br,\mathbf{0}, \omega) \cdot \bu_x}{k} &= \frac{e^{\imath kr}}{4\pi kr} \left[f_1(kr) \bu_x - f_2(kr) \sin \theta \cos\phi~ \bu_r  \right] \\
\frac{\mathbf{G}_0(\br,\mathbf{0}, \omega) \cdot \bu_y}{k} &= \frac{e^{\imath kr}}{4\pi kr} \left[f_1(kr) \bu_y - f_2(kr) \sin \theta \sin\phi~ \bu_r  \right]
\end{align*}
from which the radial component of the electric field is, in each case,
\begin{align*}
\frac{E_r^{[x]}}{P_0/\sqrt{\eps_0}} &= \frac{e^{\imath kr}}{4\pi k r} \sin \theta \cos\phi \left[ f_1(kr) - f_2 (kr) \right]  \\
\frac{E_r^{[y]}}{P_0/\sqrt{\eps_0}} &= \frac{e^{\imath kr}}{4\pi k r} \sin \theta \sin\phi \left[ f_1(kr) - f_2 (kr) \right]  ~ .
\end{align*} 
Following the same procedure, one gets the coefficients as
\begin{align*}
\Alm^{(1),[x]} = &\frac{ e^{\imath ka}}{2\imath \pi \sqrt{\ell(\ell+1)} \xi_\ell^{(1)}(ka)} \frac{-\imath ka +1}{ka} \\
& \times  \int_\Omega  \sin^2 \theta \cos\phi ~\Ylm^{*}(\theta,\phi) \d \theta \d \phi \\
\Alm^{(1),[y]} = &\frac{ e^{\imath ka}}{2\imath \pi \sqrt{\ell(\ell+1)} \xi_\ell^{(1)}(ka)} \frac{-\imath ka +1}{ka} \\
& \times \int_\Omega  \sin^2 \theta \sin\phi ~\Ylm^{*}(\theta,\phi) \d \theta \d \phi
\end{align*}
from which we develop the two integrals as
\begin{align*}
I_{\ell,m}^{[x]} = &\sqrt{\frac{2\ell+1}{4\pi}} \sqrt{\frac{(\ell-m)!}{(\ell+m)!}} \int_0^\pi \sin^2 \theta P_\ell^m(\cos \theta) ~\d \theta \\
&\times  \int_0^{2\pi}  \cos\phi e^{-\imath m \phi} ~\d \phi \\
I_{\ell,m}^{[y]} = &\sqrt{\frac{2\ell+1}{4\pi}} \sqrt{\frac{(\ell-m)!}{(\ell+m)!}} \int_0^\pi \sin^2 \theta P_\ell^m(\cos \theta) ~\d \theta\\
&\times \int_0^{2\pi}  \sin\phi e^{-\imath m \phi} ~\d \phi~ .
\end{align*}
The integrals over $\phi$ are done in (\ref{result_planarwave}). Then we consider the similar integral over $\theta$, which can be derived using orthogonality relations (\ref{ortho_Plm}) with $\ell'=1$ and $m=\pm1$. With the explicit functions $P_1^1(\cos\theta) = -\sin\theta$ and $P_1^{-1}(\cos\theta) = \sin\theta/2$, one gets that this integral is $-4/3$ for $\ell=1,m=1$, it is $2/3$ for $\ell=1,m=-1$ and zero for any other values of the indices. After replacement, one gets directly 
\begin{align*}
&I_{1,m}^{[x]} = -m\sqrt{\frac{2\pi}{3}}
&I_{1,m}^{[y]} = \imath \sqrt{\frac{2\pi}{3}} 
\end{align*}
for $m=\pm1$, and both quantities vanish for any other values of $\ell$ and $m$. This yields the same kind of simplification for the only non-zero coefficients of the spherical electric modes:
\begin{align*}
&A_{1,m}^{(1),[x]} = \frac{-m}{\sqrt{12\pi}}
&A_{1,m}^{(1),[y]} = \frac{\imath |m|}{\sqrt{12\pi}} ~ .
\end{align*}

Remarking that the coefficient of the field generated by a general dipole $\bp_0=p_0 \bu$ is $\Alm^{(1)} = \Alm^{(1),[x]} (\bu\cdot \bu_x) + \Alm^{(1),[y]} (\bu\cdot \bu_y) + \Alm^{(1),[z]} (\bu\cdot \bu_z) $, its decomposition can be written in a matrix form:
%
\begin{align}
\begin{pmatrix} \vdots \\ \Alm^{(1)} \\ \vdots \\ \hline \vdots \\ \Blm^{(1)} \\ \vdots \end{pmatrix} = 
\underbrace{\left( \begin{array}{c} \vdots \\ \Alm^{(1),[x]} \\ \vdots \\ \hline \vdots \\ \Blm^{(1),[x]} \\ \vdots \end{array} , 
\begin{array}{c}  \vdots \\ \Alm^{(1),[y]} \\ \vdots \\ \hline \vdots \\ \Blm^{(1),[y]} \\ \vdots \end{array}, 
\begin{array}{c}  \vdots \\ \Alm^{(1),[z]} \\ \vdots \\ \hline \vdots \\ \Blm^{(1),[z]} \\ \vdots \end{array} \right)}_{Q} \cdot  
\bu
\end{align} 
and the $\Blm^{(1)}$ are all zero. Finally, the operator $Q$ defined in (\ref{def_Q}) is the matrix of size $N_\sph$ by 3:
\begin{align}\label{result_Q}
Q=\frac{1}{\sqrt{12\pi}} \begin{pmatrix} 
1 & \imath  & 0 \\ 0 & 0 & \sqrt{2} \\ -1 & \imath & 0 \\ 
 0 & 0 & 0 \\ \vdots & \vdots & \vdots \\ 0 & 0 & ~ 0 ~  \end{pmatrix}
\end{align}
which only non-zero elements  are the first lines corresponding to coefficients $\Alm^{(1)}$ with $\ell=1, m\in\{-1,0,1\}$.

Let us remark that $Q$ is connected to $F$ through $Q=\imath F^\dagger$ and that $\sqrt{6\pi}Q$ is as well pseudo unitary in the sense that 
\begin{align*}
Q^\dagger Q = \frac{I}{6 \pi} 
\end{align*}
where the identity acts on $\mathds{R}^3$, while $Q Q^\dagger=I_{(A,\ell=1)}/(6\pi)$ with $I_{(A,\ell=1)}$ the identity operator for coefficients $A_{\ell,m}$ with $\ell=1, m\in \{-1,0,  1 \}$ and zero elsewhere. As a consequence one has the additional relations
\begin{align*}
&FQ = \frac{\imath I}{6 \pi}
&QF = \frac{\imath I_{(A,\ell=1)}}{6\pi} ~ .
\end{align*}

\end{document}